\documentclass[prl,twocolumn]{revtex4-1}

\usepackage[english]{babel}
\usepackage{graphicx}
\usepackage{amssymb}
\usepackage[T1]{fontenc}
\usepackage[applemac]{inputenc}
\usepackage{textcomp}
\usepackage{epsfig}
\usepackage{mathrsfs}

\def\comment#1{}
\newcommand{\Tr}{\mbox{Tr\,}}

\newcommand{\beg}{\begin{eqnarray}}
\newcommand{\eee}{\end{eqnarray}}

\def\cm#1{}

\newcommand{\be}{\begin{equation}}
\newcommand{\ee}{\end{equation}}
\newcommand{\ba}{\begin{eqnarray}}
\newcommand{\ea}{\end{eqnarray}}
\newcommand{\beq}{\begin{equation}}
\newcommand{\eeq}{\end{equation}}
\newcommand{\bea}{\begin{eqnarray}}
\newcommand{\eea}{\end{eqnarray}}
\newcommand{\bastar}{\begin{eqnarray*}}
\newcommand{\eastar}{\end{eqnarray*}}

\newcommand{\ignore}[1]{}

\begin{document}

\title{Spin-charge transformation of lattice fermion models: duality  approach for diagrammatic simulation of strongly correlated systems}
\author{Johan~Carlstr\"om}
\affiliation{
 Department of Physics, Stockholm University, 106 91 Stockholm, Sweden
}
\date{\today}
\begin{abstract}
I derive a dual description of lattice fermions, specifically focusing on the t-J and Hubbard models, that allow diagrammatic techniques to be employed efficiently in the strongly correlated regime, as well as for systems with a restricted Hilbert space.
These constructions are based on spin-charge transformation, where the lattice fermions of the original model are mapped onto spins and spin-less fermions. This mapping can then be combined with Popov-Fedotov fermionisation, where the spins are mapped onto lattice fermions with imaginary chemical potential. 
The resulting models do not contain any large expansion parameters, even for strongly correlated systems. Also, they exhibit dramatically smaller corrections to the density matrix from nonlinear terms in the Hamiltonian.  
The combination of these two properties means that they can be addressed with diagrammatic methods, including simulation techniques based on stochastic sampling of diagrammatic expansions.   
\end{abstract}
\maketitle

\subsection{Introduction}
The Hubbard model \cite{Hubbard238}, and the closely related t-J model \cite{PhysRevB.18.3453} occupy a central position in condensed matter physics. Though conceived more broadly as descriptions of strongly correlated systems they are now perhaps most famous as paradigmatic descriptions of the high-temperature superconductors. 
As such, they have motivated a wide range of theoretical works, and even prompted experimental realisations in ultra cold atomic systems. 

Important theoretical progress has been made with a range approximative techniques, ranging from Dynamic Mean Field Theory to Projected Wave Functions and Gutzwiller Wave Function methods. Central aims of these works are the pairing mechanism, pairing symmetry, the parameter range that exhibits superconductivity and the extent to which they reproduce the physics observed in the cuprates \cite{PhysRevLett.62.324,PhysRevB.75.045118,PhysRevB.77.033101,PhysRevLett.106.047004,PhysRevLett.110.216405,PhysRevLett.87.217002,PhysRevB.95.024506,PhysRevB.37.533}.

Despite considerable advances, a reliable phase diagram for the Hubbard and t-J models has not been obtained, and in many cases different numerical protocols display notable discrepancies \cite{2006cond.mat.10710S}. Ultimately, numerical treatment of strongly correlated systems face a number of challenges: The possibility of phase separation, stripe formation \cite{Nature375-561}, or other types of spatial ordering \cite{GranularCuprates} is highly problematic for any methods based on finite cluster descriptions, as the domain size and shape, as well as boundary conditions can affect the outcome \cite{2006cond.mat.10710S}. 
Meanwhile, the prospect of competing states that are narrowly spaced in free energy makes uncontrolled approximations potentially very misleading \cite{Dagotto257,2006cond.mat.10710S}.

Recently, there has been a rapid development of computational methods for addressing interacting fermionic many-body systems. 
A few examples of techniques being explored in this capacity are DMRG \cite{PhysRevLett.69.2863}, DCA \cite{RevModPhys.77.1027}, Linked cluster expansion \cite{PhysRevE.89.063301,PhysRevE.75.061119}, Auxiliary-field quantum Monte Carlo \cite{afqmc} and Diagrammatic Monte Carlo \cite{PhysRevLett.99.250201,0295-5075-110-5-57001}.  
These promising developments have not only sparked a renewed interest in correlated electrons, but have even resulted in the formation of a new scientific body, namely the Simons collaboration on the many-electron problem \cite{simon}. 
Recently, this group published a benchmark, comparing a wide range of computational techniques when applied to the Hubbard model. 

The conclusion in this work is that a substantial part of the ground state properties are now under control, and that these appear to connect smoothly to finite temperature results. The most substantial uncertainties exist near half-filling and for interaction strengths of $4\le U/t \le 8$ \cite{PhysRevX.5.041041}.

Given the challenges posed by high-temperature superconductivity, and strongly correlated systems in general, methodological development is vital, and in particular, controllable numerical techniques is a key factor in obtaining reliable phase diagrams of the relevant models. 

Recently, part of the phase diagram for the Hubbard model was obtained using Diagrammatic Monte Carlo \cite{0295-5075-110-5-57001}.
This technique is based on stochastic sampling of the diagrammatic expansion which can be conducted directly in the macroscopic limit, and correspondingly, it is not susceptible to inappropriate boundary conditions or domain shape. Also, it does not employ any approximations beyond truncation of the diagrammatic series, meaning that results are ultimately controllable. 
With this method, the principal obstacle to simulation of strongly correlated systems is the rate of convergence-- Strong inter-particle interaction implies a diagrammatic series with a large expansion parameter, and this is generally detrimental to convergence. Also, in these regimes the correction to the density matrix due to interactions can grow extremely large, which also hinders convergence. 
Thus, the work \cite{0295-5075-110-5-57001} was limited to moderate onsite repulsions of $U/t\le 4$ and filling factors up to $\sim 70\%$. Crucially however, it does report controllable and highly accurate results for the pairing symmetry.

The fact that diagrammatic methods are limited to fermionic systems and moderate expansion parameters has motivated a search for analytical techniques that allow mapping of various many-body problems onto descriptions that comply with these restrictions. An example of this is Popov-Fedotov fermionisation, which allows spin-models to be mapped onto lattice fermions with imaginary chemical potential \cite{JETP.67.535}, and which has successfully been used in combination with diagrammatic techniques to address frustrated spin models \cite{PhysRevB.87.024407,PhysRevLett.110.070601,PhysRevLett.116.177203}. A more recent development is universal fermionisation, which has been proposed as a means to remove large or infinite interactions terms from lattice fermion models \cite{PhysRevB.84.073102}.

This article will discuss transformation techniques aimed at the strongly correlated regime. In particular it will be shown that it is possible to construct mappings of the Hubbard and t-J models based on spin-charge transformation in combination with fermionisation. Thus transformed models are free of large expansion parameters, and also exhibit much smaller corrections to the density matrix from interaction terms. These two properties are essential for treatment with diagrammatic Monte Carlo.

\subsection{Fermionisation}

The fermionisation technique proposed by Popov and Fedotov \cite{JETP.67.535} allows spins to be mapped onto lattice fermions, and is very simple. The central idea is that a spin state $\sigma$ maps onto a state that has one spin-$\sigma$ fermion according to
\bea
|\sigma\rangle \to |n_\sigma=1,n_{\bar{\sigma}}=0\rangle,
\eea
where $\bar{\sigma}=-\sigma$. The terms in the Hamiltonian that operate on the spins must then be expressed in fermionic operators accordingly. This construction does however introduce unphysical states that corresponds to either $0$ or $2$ fermions that have no corresponding spin-state. Remarkably, these can be projected out from the partition function by the introduction of an imaginary chemical potential term of the form:
\bea
H'= H+\sum_j\frac{i\pi}{2\beta}(n_{j \uparrow}+n_{j \downarrow}-1).\label{Popov}
\eea
This term removes unphysical contributions to the trace by ascribing to them complex phases such that they cancel. Once the spin-operators in $H$ are reformulated as fermionic operators, the problem can be treated with Diagrammatic Monte Carlo \cite{PhysRevB.87.024407,PhysRevLett.110.070601,PhysRevLett.116.177203}.  It should be noted that the Hamiltonian (\ref{Popov}) is not Hermitian. However, the non-Hermitian part only acts on unphysical states since it is zero in the physical subspace. Thus, $H'$ behaves as a symmetric operator in the physical subspace.

A more general fermionisation technique aimed at strongly correlated systems is universal fermionisation \cite{PhysRevB.84.073102}. This method allows the encoding of restricted Hilbert spaces by the introduction of projection terms in the Hamiltonian: 
Assuming a theory $H$ on an unrestricted Hilbert space we may construct the corresponding theory in a restricted Hilbert space $H'$ as follows:
\bea
H'=H+\sum_j \frac{i\pi}{\beta}P_j (n_{j A}-1/2), \; [H,P_j]=0\label{universal}.
\eea
Here, $P_j$ is a projection operator onto the forbidden subspace (i.e. it returns $1$ if the site $j$ has a forbidden configuration and zero otherwise).
The term $n_A=c^\dagger_A c_A$ is a number operator for a field of auxiliary spin-less fermions, introduced in addition to the original degrees of freedom. 
By this construction, contributions to the partition function from forbidden states obtain a complex phase that results in their cancelation. Thus for example, summing over the contributions that have $P_j=1$, we obtain cancelation between terms corresponding to $n_{A,j}=0,1$ as they enter with a phase difference of $\pi$ in the trace. 
In the allowed sector, where $P_j=0$, the introduction of an additional fermionic field simply amounts to multiplying the trace with a trivial number ($2^N$, where N is the number of sites).  As in (\ref{Popov}), the non-Hermitian term operates only on states outside the physical subspace.

This technique can for example be employed to the t-J model to remove ``forbidden'' doubly occupied sites from the trace. It has also been proposed as a way of dealing with the Hubbard model at large $U$, by removing doubly occupied sites and then reintroducing them in the form of bosons that are subsequently fermionised, a procedure referred to as second fermionisation \cite{PhysRevB.84.073102}. 

Although universal fermionisation can encode restrictions in the Hilbert space for the t-J model, and also remove large onsite repulsion in the Hubbard model, diagrammatic treatment of these systems breaks down at half filling. This in turn is related to increasingly large corrections to the density matrix from nonlinear terms in this limit. This effect can be quantified by considering a restricted Hilbert space in the atomic limit and computing an expansion in the projection operator. Let $H=-\mu (n_{\uparrow}+n_{\downarrow} )$ and introduce a parameter $\xi$ for the projection terms so that $\xi=1$ corresponds to the fully projected theory. We then find:
\bea
H' =-\mu (n_{\uparrow}+n_{\downarrow} )+\xi  \frac{i\pi }{\beta}n_{\uparrow}n_{\downarrow}(n_{A}-1/2),\\
\langle n\rangle=\frac{2 e^{\beta \mu}+2e^{2\beta \mu} \cos(\pi \xi/2)}{1 +2 e^{\beta \mu}+e^{2\beta\mu} \cos(\pi \xi/2)}. \label{atomic}
\eea 
A diagrammatic expansion in the projection operator is equivalent to an expansion in $\xi$, thus one can easily determine the perturbative order required to obtain $\langle n\rangle$ with a given accuracy goal. With the (very forgiving) requirement that the error may not exceed $1\%$ we find the required expansion order to be as follows: For the case $\beta\mu=0$ and $\langle n\rangle$=2/3 we require 4 orders, for the case $\beta\mu=3/2$ and $\langle n\rangle\approx 0.9$ we require 16 orders, and for $\beta\mu=4$ and $\langle n\rangle\approx 0.99$ we find that an expansion to order 100 is not sufficient. To put this in perspective, with state of the art Diagrammatic Monte Carlo, expansion orders up to $6\le n_{max}\le8$ are typically viable for fermionic field theories, so a wide range of filling factors relevant to the cuprates are simply not tractable with this approach. 

The increasingly poor convergence properties at small doping are related to large corrections to the density matrix in the following way: Let $Z_0=Z(\xi=0)$ and $Z_1=Z(\xi=1)$ be the partition functions of the unprojected and projected theories. If $\beta\mu$ is large and positive, then $Z_0$ is dominated by the ``forbidden'' state with double occupation, implying that $Z_1\sim e^{-\beta\mu} Z_0$ for $\beta\mu\gg 1$. Thus, projection changes the weight $w$ of an allowed state by
\bea
w_0=\frac{e^{E}}{Z_0}\to w_1= \frac{e^{E}}{Z_1}\sim   \frac{e^{E}}{Z_0}e^{\beta\mu},
\eea
which is singular at half filling when $\beta\mu\to\infty$. 

To address this problem, it is not sufficient to transform interaction terms, as this would preserve the corrections to the density matrix. Rather, it is necessary to change the basic degrees of freedom in such a way that the new theory is devoid of any singularities or large expansion parameters. 

\subsection{Spin-charge transformation}
It is possible to remove the singular corrections to the density matrix by mapping onto a dual description that treats holes and doubly occupied sites (doublons) on an equal footing. 
In the original models the Hilbert space is spanned by
\bea
|0\rangle,\;|\uparrow\rangle,\; |\downarrow\rangle, \; |\downarrow\uparrow\rangle.
\eea
This can be mapped onto a description in terms of spins and spin-less fermions according to
 \bea\nonumber
|\downarrow\uparrow\rangle  \to |n_\Delta=1\rangle\times |D\rangle,\;\;\;
|0\rangle  \to |n_\Delta=1\rangle\times |H\rangle,\;\\
\;|\uparrow\rangle\to |n_\Delta=0\rangle\times |\uparrow\rangle,\;\;\;
|\downarrow\rangle \to   |n_\Delta=0\rangle \times |\downarrow\rangle.\;\label{statemap}
\eea
In this representation, a basis vector is characterised by an occupation number for the spin-less fermions $n_\Delta=0,1$ and a spin that takes values $\uparrow,\;\downarrow$.
The case $n_\Delta=1$ corresponds to either a hole or a doublon, depending on the corresponding spin state. In principe we can allow $|D\rangle $ and $ |H\rangle,$ to take any values on the $SU(2)$ sphere, provided that they are orthogonal, $\langle H|D\rangle=0$, in order for Eq. (\ref{statemap}) to define an orthogonal basis.
However in practice the simplest choice is to take the eigenbasis of $S_z$, for example 
$|D\rangle =|\uparrow\rangle,\;|H\rangle =|\downarrow\rangle.$

A mapping of the original fermionic operators that preserves the anti-commutation relations can be constructed according to 
\bea\nonumber
c_{i\sigma}^\dagger \to \Delta_i(|\sigma,i\rangle\langle H,i|)+\sigma\Delta_i^\dagger(|D,i\rangle\langle \bar{\sigma},i|)\\
c_{i\sigma} \to \Delta_i^\dagger(|H,i\rangle\langle \sigma,i|)+\sigma\Delta_i(|\bar{\sigma},i\rangle\langle D,i|)\label{SCT}
\eea
where $\sigma=\pm 1$, $|\sigma=1\rangle=|\uparrow\rangle$, $|\sigma=-1\rangle=|\downarrow\rangle$, and $\Delta_i,\; \Delta_i^\dagger$ are annihilation and creation operators of a spin-less fermion on the site $i$. The operator $|\sigma,i\rangle \langle \sigma',i|$ acts on the spin of the lattice site $i$, where it changes the value from $\sigma'$ to $\sigma$. Finally, $\bar{\sigma}=-\sigma$.
A natural comparison to this construction is the slave-Boson approach \cite{0305-4608-6-7-018,PhysRevLett.57.1362}. However, the introduction of a spin, rather than a Boson results in a direct correspondence between states in the respective representations (\ref{statemap}), thus eliminating the need for any restrictions on the Hilbert space. This greatly simplifies diagrammatic treatment. 
 
\subsection{Hubbard model}
The Hubbard model takes the form
\bea
H=-\sum_{ i,j,\sigma} t_{ij}c^\dagger_{i\sigma}c_{j\sigma}+\sum_i \Big(U n_{i\uparrow}n_{i\downarrow}-\mu n_i\Big),\label{Hubbard0}
\eea
where $c^\dagger_\sigma,\; c_\sigma$ are creation and annihilation operators of a spin-$\sigma$ fermion on $i$, $n_{i,\sigma}=c_{i,\sigma}^\dagger c_{i,\sigma}$ and $n_i=n_{i,\uparrow}+n_{i,\downarrow}$. 
To obtain the spin-charge transformed Hubbard model we insert \ref{SCT} into \ref{Hubbard0}.
The inter-site terms then map according to 
\bea\nonumber
 c_{i\sigma}^\dagger c_{j\sigma}\to
\big\{\Delta_i|\sigma,i\rangle \langle H,i|+\sigma \Delta_i^\dagger |D,i\rangle \langle \bar{\sigma},i|\big\}\\ \nonumber\times\big\{\Delta_j^\dagger|\sigma,j\rangle \langle H,j|+\sigma \Delta_j |D,j\rangle \langle \bar{\sigma},j|\big\}\\
=-\Delta_j^\dagger \Delta_i |\sigma,i\rangle\langle H,i|\times |H,j\rangle \langle \sigma,j| \label{holeprop}\\
+\Delta_i^\dagger\Delta_j |D,i\rangle\langle \bar{\sigma},i| \times |\bar{\sigma},j\rangle \langle D,j| \label{doubprop}\\
+\sigma \Delta_i\Delta_j |\sigma,i\rangle \langle H,i| \times |\bar{\sigma},j\rangle\langle D,j|\label{annihilation}\\
+\sigma  \Delta_i^\dagger \Delta_j^\dagger |D,i\rangle \langle \bar{\sigma},i |\times | H,j\rangle \langle \sigma,j|.\label{creation}
\eea
The corresponding physical processes are: Hole propagation (\ref{holeprop}), Doublon propagation (\ref{doubprop}), Annihilation of a doublon-hole pair (\ref{annihilation}) and Creation of a doublon-hole pair (\ref{creation}).

The single-site terms in the Hubbard model map according to
\bea
H_{j}=-\mu \sum_\sigma c_{j\sigma}^\dagger c_{j\sigma} + U c_{j\uparrow}^\dagger c_{j\uparrow}c_{j\downarrow}^\dagger c_{j\downarrow}\to\\
-\mu (1-\Delta_j^\dagger\Delta_j) +(U-2\mu)\Delta_j^\dagger \Delta_j |D,j\rangle \langle D,j|,
\eea
where the constant term $-\mu$ can be dropped. 

As spins can not be treated directly with diagrammatic techniques, they have to be mapped onto fermions via Popov-Fedotov fermionisation \cite{JETP.67.535}. Let the state vectors map according to
\bea
|S=\sigma\rangle \to |n_\sigma=1,n_{\bar{\sigma}}=0\rangle,
\eea
and define the creation and annihilation operators of the 'spin-fermions' $a_{i,\sigma}^\dagger,a_{i,\sigma}$. This construction introduces unphysical states that with $n_\sigma=n_{\bar{\sigma}}=0$ and $n_\sigma=n_{\bar{\sigma}}=1$ that have no spin counterpart, and these have to be projected out through an imaginary chemical potential term of the form (\ref{Popov}).

The spin operators map according to 
\bea\nonumber
|\sigma\rangle \langle \sigma| \to \Big(\frac{1}{2}+\sigma S_z\Big),\;S_z=\frac{1}{2}\big(a^\dagger_\uparrow a_\uparrow-a^\dagger_{\downarrow} a_{\downarrow} \big)\\
|\sigma\rangle \langle \bar{\sigma}| \to  a^\dagger_\sigma a_{\bar{\sigma}}.\label{spinop}
\eea
This construction ensures that in the physical sector we have $|\sigma\rangle \langle \sigma'|=a^\dagger_\sigma a_{\sigma'}$.
Meanwhile, the unphysical states form a degenerate eigenspace of the spin operators:
\bea\nonumber
|\sigma\rangle\langle \sigma| a^\dagger_\uparrow a^\dagger_\downarrow|0\rangle=\frac{1}{2} a^\dagger_\uparrow a^\dagger_\downarrow|0\rangle,\;
|\sigma\rangle\langle \sigma   |0\rangle=\frac{1}{2} |0\rangle
\\
|\sigma\rangle\langle \bar{\sigma}| a^\dagger_\uparrow a^\dagger_\downarrow|0\rangle=0,\;
|\sigma\rangle\langle \bar{\sigma}  |0\rangle=0.
\eea
Since the spin operators commute with $a^\dagger_\uparrow a^\dagger_\downarrow$, so does $H$. 
However, $H'$ does not commute with this operator due to the imaginary chemical potential term (\ref{Popov}). In particular we find
\bea
\Big[\frac{i\pi}{2\beta}(n_{j\uparrow }+n_{j\downarrow }-1),a^\dagger_{m\uparrow } a^\dagger_{m\downarrow }\Big]=\frac{i\pi\delta_{j m}}{\beta}a^\dagger_{m\uparrow } a^\dagger_{m\downarrow }.
\eea
Defining the self-adjoint projection operator onto the empty site $p_m=(1-n_{m\uparrow})(1-n_{m\downarrow})$ and taking $D_m^\dagger=a^\dagger_{m\uparrow} a^\dagger_{m\downarrow}$,
we can write the trace over all states that have an unphysical occupation number at the site $m$ as
\bea
\Tr p_m(1+D_m)e^{-\beta H'} (1+D^\dagger_m) p_m \\
=\Tr p_m(1+D_m)(1+D^\dagger_m e^{-i\pi})e^{-\beta H'}  p_m\\
=\Tr p_m(1-D_mD_m^\dagger) p_m e^{-\beta H'}=0
\eea
and so we find that contributions from unphysical states to the trace vanish as required.  
The spin-charge transformed and fermionised Hubbard model then takes the form
\bea
H\nonumber
=\sum_{ij,\sigma}t_{ij}\Big\{\Delta_j^\dagger \Delta_i |\sigma,i\rangle\langle H,i|\times |H,j\rangle \langle \sigma,j| \\ \nonumber
-\Delta_i^\dagger\Delta_j |D,i\rangle\langle \bar{\sigma},i| \times |\bar{\sigma},j\rangle \langle D,j| 
\\ \nonumber
-\sigma \Delta_i\Delta_j |\sigma,i\rangle \langle H,i| \times |\bar{\sigma},j\rangle\langle D,j|
\\\nonumber
-\sigma  \Delta_i^\dagger \Delta_j^\dagger |D,i\rangle \langle \bar{\sigma},i |\times | H,j\rangle \langle \sigma,j|\Big\}
\\\nonumber
+\sum_{i}\Big\{\mu \Delta_i^\dagger\Delta_i +(U-2\mu)\Delta_i^\dagger \Delta_i |D,i\rangle \langle D,i|\\
+\frac{ i\pi}{2\beta}(n_{i\sigma}+n_{i\bar{\sigma}}-1)\Big\}\label{SCT_Hubbard}
\eea
where we have defined
\bea
|\sigma,i\rangle \langle \sigma,i| = \Big(\frac{1}{2}+\sigma S_{i,z}\Big),\;|\sigma,i\rangle \langle \bar{\sigma},i| =  a^\dagger_{i,\sigma} a_{i,\bar{\sigma}}.\label{FP}
\eea
The price of this transformation is a higher degree of complexity of the Hamiltonian, but we also find that the onsite repulsion $U n_{i\uparrow}n_{i\downarrow}$ is replaced by a much weaker interaction $\sim U-2\mu$ which describes the energy difference between holes and doublons. More important however is that it is free of singular corrections to the density matrix at half filling. Thus, unlike the original Hubbard model, this theory can be transformed further via universal fermionisation if necessary. For instance, repeating the treatment in the atomic limit (\ref{atomic}) we find
\bea
H' =\mu \Delta^\dagger \Delta+\xi  \frac{i\pi }{\beta} P(n_{A}-\frac{1}{2})  +\frac{ i\pi}{2\beta}(n_{\uparrow}+n_{\downarrow}-1)\;\;\;\;\;\;\\
\langle 1-\Delta^\dagger\Delta \rangle=1-\frac{e^{-\beta \mu}(1+ \cos(\pi \xi/2))}{2+e^{-\beta \mu}(1+ \cos(\pi \xi/2))} =\;\;\;\;\;\\
\frac{1}{1+e^{-\beta\mu}}+\frac{e^{\beta \mu} \pi^2\xi^2}{16(1+e^{\beta\mu})^2} - \frac{e^{\beta\mu}(e^{\beta\mu}-2)\pi^4\xi^4 }{768(1-e^{\beta\mu})^3}+...\;\;\;\;\;\;
\eea
and so when $\beta\mu\gg 1$, the corrections to the particle density $\langle 1-\Delta^\dagger\Delta\rangle$ are suppressed as $\sim e^{-\beta \mu}$ at all orders finite in $\xi$, leading to exceptionally fast convergence.  

The best way to solve (\ref{SCT_Hubbard}) depends on model parameters. 
If $U-2\mu$ is of order unity, then it can be solved directly through diagrammatic expansion. 
If $U\to \infty$, then doublons can be regarded as a forbidden state and removed through universal fermionisation along with creation/annihilation of doublon-hole pairs and doublon propagation. The problem can then be treated with diagrammatic methods. 
If $U\gg t$ but still finite, then this problem can under certain circumstances be approximated by the t-J model (which will be addressed below).
The most challenging situation arises when $U-2\mu$ is large, but $U/t$ is not sufficiently large to motivate treatment with the t-J model. In this case it is necessary to employ second fermionisation \cite{PhysRevB.84.073102}. This involves removing the doubly occupied sites via universal fermionisation and then reintroducing them as hardcore bosons which are subsequently fermionised. The resulting model will be free of large expansion parameters and can be solved through diagrammatic expansion. 

\subsection{t-J model}
When the filling factor is less than unity and the onsite repulsion is much greater than the bandwidth, the t-J model can be derived as an effective description of the Hubbard model through canonical transformation of the latter and truncation of terms subleading to $\sim t^2/U$ \cite{PhysRevB.18.3453}. This model takes the form
\bea\nonumber
H=\sum_{ ij,\sigma}t_{ij} (1-c^\dagger_{i\bar{\sigma}}c_{i\bar{\sigma}}) c^\dagger_{i\sigma} c_{j\sigma}  (1-c^\dagger_{j\bar{\sigma}}c_{j\bar{\sigma}})
\\
+\sum_{ ij}J_{ij}\; (S_i\cdot S_j-\frac{n_i n_j}{4})
-\sum_j \mu n_{j}, \label{classicTJ}
\eea
where $n_i=c^\dagger_{i\uparrow}c_{i\uparrow}+c^\dagger_{i\downarrow}c_{i\downarrow}$ is the number operator.

Inserting (\ref{SCT}) into (\ref{classicTJ}) and fermionising the spins, we obtain the transformed t-J model:
\bea\nonumber
H=\sum_{ij,\sigma}t_{ij}\Delta_j^\dagger \Delta_i |\sigma,i\rangle\langle H,i|\times |H,j\rangle \langle \sigma,j| \\\nonumber
+\sum_{ij} J_{ij} (1-\Delta_i^\dagger  \Delta_i)(1-\Delta_j^\dagger\Delta_j) \Big(S_i \cdot S_j-\frac{1}{4}\Big)\\
+\sum_i \Big\{ \mu \Delta_i^\dagger\Delta_i 
+\frac{ i\pi}{2\beta}(n_{i\sigma}+n_{i\bar{\sigma}}-1)
\Big\},\label{tJ}
\eea
where the spin operators are as before given by (\ref{FP}). 
Since doubly occupied sites are forbidden they have to be projected out via universal fermionisation. The projection operator ($P_j$ in Eq. \ref{universal}) maps according to
\bea
P_j=n_{j\uparrow}n_{j\downarrow}\to \Delta_j^\dagger  \Delta_j |D,j\rangle \langle D,j|
\eea
and we find that the projected t-J model takes the form
\bea
H'=H+\frac{i\pi}{\beta}\sum_j  \Delta^\dagger_j \Delta_j |D,j\rangle\langle D,j| (c_{jA}^\dagger c_{jA}-1/2),\label{tJF}\;\;\;\;\;\;
\eea
where $c_{jA}^\dagger, \; c_{jA}$ are creation and annihilation operators of the auxiliary fermionic field. 

As doubly occupied sites are not part of the physical space of the t-J model, the spin-less fermions can unambiguously be interpreted as holes with a chemical potential $-\mu$. When $\beta\mu$ is large and $\mu \gg zt$ the hole density vanishes, and we recover the fermionised Heisenberg model. Thus, the transformed t-J model can be regarded as a generalisation of the Popov-Fedotov technique to the case of a doped system.

In conclusion, we find that by applying spin-charge transformation to fermionic lattice models with large or infinite onsite repulsion, we can remove the singular corrections to the density matrix at half filling. This opens up the way for fermionisation techniques to be used in this parameter region to encode restricted Hilbert spaces or transform strong inter-particle interactions via second fermionisation. 
Thus transformed models posses expansion parameters of order unity and can be efficiently addressed with diagrammatic simulation techniques. 
These methods have proven to be extremely accurate for a wide range models with moderate interaction terms. They also give access to a wide range of observables, including density of states, particle/spin correlators and effective dispersion relations.  
Through the proper transformation techniques, they can also be applied in the strongly correlated regime.

This work was supported by the Wenner-Gren Foundations in Stockholm and 
the Simons Collaboration on the Many Electron Problem.
The author would like to acknowledge vital input and advice from Boris Svistunov and Nikolay Prokof'ev.

\bibliography{biblio}

\end{document}